\documentclass[11pt,a4paper]{article}
\usepackage{jheppub}
\usepackage{graphicx}
\usepackage{amssymb,amsfonts,amsmath}
\usepackage{epsfig}
\def\lsim{\mathrel{\rlap{\lower3pt\hbox{\hskip1pt$\sim$}}
\raise1pt\hbox{$<$}}}
\def\gsim{\mathrel{\rlap{\lower3pt\hbox{\hskip1pt$\sim$}}
\raise1pt\hbox{$>$}}}
\newcommand\dslash{\slash\hspace*{-0.7em}D}
\newcommand\avr[1]{\left\langle{#1}\right\rangle}
\newcommand{\beq}{\begin{equation}}
\newcommand{\eeq}{\end{equation}}
\newcommand{\bea}{\begin{eqnarray}}
\newcommand{\eea}{\end{eqnarray}}

\newcommand{\etal}{, \textit{et. al.~~}}
\title{Fluctuations of conserved charges at finite temperature from lattice QCD}
\author[a]{Szabolcs Bors\'anyi}
\author[a,b,c]{Zolt\'an Fodor}
\author[b]{S\'andor D. Katz}
\author[a,c]{Stefan Krieg}
\author[d,e]{Claudia Ratti}
\author[a]{K\'alman Szab\'o}
\affiliation[a]{
Dept. of Physics, Wuppertal University,\\
Gaussstr. 20, D-42119 Wuppertal, Germany}
\affiliation[b]{
Inst. for Theoretical Physics, E\"otv\"os University,\\
P\'azm\'any 1, H-1117 Budapest, Hungary}
\affiliation[c]{
J\"ulich Supercomputing Centre,\\
Forschungszentrum J\"ulich, D-52425}
\affiliation[d]{
Dip. di Fisica Teorica, Universit\`a di Torino,\\
via Giuria 1, I-10125 Torino, Italy
}
\affiliation[e]{
INFN, Sezione di Torino
}

\emailAdd{borsanyi@uni-wuppertal.de}

\abstract{
We present the full results of the Wuppertal-Budapest lattice QCD collaboration
on flavor diagonal and non-diagonal quark number susceptibilities with 2+1
staggered quark flavors, in a temperature range between 125 and 400 MeV.
The light and strange quark masses are set to their physical values.
Lattices with $N_t=6,~8,~10,~12,~16$ are used. We perform a continuum
extrapolation of all observables under study.  A Symanzik improved gauge and a
stout-link improved staggered fermion action is utilized.  All results are
compared to the Hadron Resonance Gas model predictions: good agreement is found
in the temperature region below the transition.
}

\begin{document}
\maketitle

\section{Introduction}

The QCD transition, once occured in the Early Universe, is being routinely
reproduced in the laboratory, in the ultrarelativistic heavy ion collision
experiments at CERN SPS, RHIC at Brookhaven National Laboratory, ALICE at the
LHC and the future FAIR at the GSI. The most important known qualitative
feature of this transition is its cross-over nature at vanishing baryo-chemical
potential \cite{Aoki:2006we}.
A lot of effort has been invested, both theoretically and experimentally, in
order to find observables which can unambiguously signal the transition.  As
expected in a cross-over, observables follow a smooth behaviour over the
transition. The characteristic temperature of the transition depends on how
one defines it. For the renormalized chiral condensate the
Wuppertal-Budapest collaboration predicted a value around 150 MeV, which was
recently confirmed by hotQCD (see for their journal publication
\cite{arXiv:1111.1710}).

Correlations and fluctuations of conserved charges have been proposed long ago
to signal the transition \cite{Jeon:2000wg,Asakawa:2000wh}. The idea is that these
quantum numbers have a very different value in a confined and deconfined
system, and measuring them in the laboratory would allow to distinguish between
the two phases.

Fluctuations of conserved charges can be obtained as linear combinations of
diagonal and non-diagonal quark number susceptibilities, which can be
calculated on the lattice at zero chemical potential
\cite{Gottlieb:1988cq,Gavai:1989ce}. These observables can give us an insight
on the nature of the matter under study \cite{Gottlieb:1988cq,McLerran:1987pz}.
Diagonal susceptibilities measure the response of the quark number density to
changes in the chemical potential, and show a rapid rise in the crossover region.
At high temperatures they are expected to be large, if 
the quark mass is small in comparison to the temperature. At very high
temperatures diagonal susceptibilities are expected to 
approach the ideal gas limit. On the other hand, in the
low-temperature phase they are expected to be small since quarks are confined
and the only states with nonzero quark number have large masses. Agreement with
the Hadron Resonance Gas (HRG) model predictions is expected in this phase
\cite{hrg}. Non-diagonal susceptibilities give us information about the
correlation between different flavors. They are supposed to vanish in a
non-interacting quark-gluon plasma (QGP). In the approximately self-consistent
resummation scheme of hard thermal and dense loops Ref.~\cite{Blaizot:2001vr}
shows nonzero correlations between different flavors at large temperatures
due to the presence of flavor-mixing diagrams. A quantitative analysis of this
observable allows one to draw conclusions about the presence of bound states in
the QGP \cite{Ratti:2011au}.  Another observable which was proposed to this
purpose, and which can be obtained from a combination of diagonal and
non-diagonal quark number susceptibilities, is the baryon-strangeness
correlator \cite{Koch:2005vg}.

Several results exist in the literature about the study of quark number
susceptibilities on the lattice both for 2 \cite{Ejiri:2005wq} and 2+1
\cite{hotQCD2} quark flavors. However, for the first time in this paper the
susceptibilities are calculated for physical values of the quark masses and a
continuum extrapolation is performed not only for strange quark
susceptibilities \cite{Borsanyi:2010bp} but also for the light quark and the
non-diagonal ones.  We present full results of our collaboration for several of
these observables, with 2+1 staggered quark flavors, in a temperature range
between 125 and 400 MeV.  The light and strange quark masses are set to their
physical values. Lattices with $N_t=6,~8,~10,~12,~16$ are used. Continuum
extrapolations are performed for all observables under study. We compare our
results to the predictions of the HRG model with resonances up to 2.5 GeV mass
at small temperatures, and of the Hard Thermal Loop (HTL) resummation scheme at
large temperatures, when available.

\section{Observables under study} 

The baryon number $B$, strangeness $S$ and electric charge $Q$ fluctuations can be obtained, at vanishing chemical potentials, from the QCD partition function. The relationships between the quark chemical potentials and those of the conserved charges are as follows:
\bea
\mu_u&=&\frac13\mu_B+\frac23\mu_Q;
\nonumber\\
\mu_d&=&\frac13\mu_B-\frac13\mu_Q;
\nonumber\\
\mu_s&=&\frac13\mu_B-\frac13\mu_Q-\mu_S.
\label{chem}
\eea
Here the small indices $u$, $d$ and $s$ refer to up, down and strange quark
numbers, which, too,  are conserved charges in QCD. The negativ sign between
$\mu_s$ and $\mu_S$ reflects the $-1$ strangeness quantum number of the strange
quark.

Starting from the QCD pressure,
\bea
\frac{p}{T^4}=\frac{1}{VT^3}\ln Z(V,T,\mu_B,\mu_S,\mu_Q)
\eea
we can define the moments of charge fluctuations as follows:
\bea
\chi_{lmn}^{BSQ}=\frac{\partial^{\,l+m+n}p/T^4}{\partial(\mu_{B}/T)^{l}\partial(\mu_{S}/T)^{m}\partial(\mu_{Q}/T)^{n}}.
\eea
In the present paper we will concentrate on the quadratic fluctuations,
thus $l+m+n=2$. In terms of quark numbers ($N_X$) our observables read\footnote{
For simplicity we inculde the normalization $1/T^2$ in the definition of
$\chi_2^X$ and $\chi_{11}^{XY}$. In Refs.~\cite{6,7,Borsanyi:2010bp} we used
the notation $\chi_2^X/T^2$ for the same observable.
}:
\bea
\chi_{2}^{X}=\frac{1}{VT^3}\langle N_{X}^{2}\rangle
\eea
and on the correlators among different charges or quark flavors:
\bea
\chi_{11}^{XY}=\frac{1}{VT^3}\langle N_XN_Y\rangle,
\eea
where $X$ and $Y$ are one of $u$, $d$ and $s$.
Given the relationships between chemical potentials (\ref{chem}) the diagonal susceptibilities of the conserved charges can be obtained from quark number susceptibilities in the following way:
\begin{eqnarray}
\chi_2^{B}&=& \frac19 \left[
\chi_2^{u}+\chi_2^{d}+\chi_2^{s}
+2\chi_{11}^{us}+2\chi_{11}^{ds}+2\chi_{11}^{ud}\right]\,,
\nonumber\\
\chi_2^{Q}&=& \frac19 \left[
4\chi_2^u+\chi_2^d+\chi_2^s
-4\chi_{11}^{us}+2\chi_{11}^{ds}-4\chi_{11}^{ud}
\right]\,,
\nonumber\\
\chi_2^{I}&=& \frac14 \left[
\chi_2^u+\chi_2^d-2\chi_{11}^{ud}
\right]\,,
\nonumber\\
\chi_2^{S}&=& \chi_2^s\,.
\end{eqnarray}

If we do not wish to take further derivatives, we can take all three chemical potentials ($u,d,s$) to zero.
In this case, for our 2+1 flavor framework nothing distinguishes between the
$u$ and $d$ derivative: this gives slightly simplified formulae:

\begin{eqnarray}
\chi_2^{B}&=&\frac19 \left[
2\chi_2^u+\chi_2^s
+4\chi_{11}^{us}+2\chi_{11}^{ud}
\right],
\nonumber\\
\chi_2^{Q}&=& \frac19 \left[
5\chi_2^u+\chi_2^s
-2\chi_{11}^{us}-4\chi_{11}^{ud}
\right]\,,
\nonumber\\
\chi_2^{I}&=& \frac12 \left[
\chi_2^u-\chi_{11}^{ud}
\right]\,.
\label{chiB}
\end{eqnarray}

The baryon-strangeness correlator, which was proposed in Ref. \cite{Koch:2005vg} as a diagnostic to understand the nature of the degrees of freedom in the QGP, has the following expression in terms of quark number susceptibilities:

\bea
C_{BS}=-3\frac{\langle N_BN_S\rangle}{\langle N_{S}^2\rangle}=1+\frac{\chi_{11}^{us}+\chi_{11}^{ds}}{\chi_2^s}.
\label{cbs}
\eea


\section{Details of the lattice simulations \label{lattice}}

\subsection{The lattice action} 
The lattice action is the same as we used in \cite{6, 7}, namely a tree-level
Symanzik improved gauge, and a stout-improved staggered fermionic action (see
Ref. \cite{Aoki:2005vt} for details).  The stout-smearing
\cite{Morningstar:2003gk} yields an improved discretization of the
fermion-gauge vertex and reduces a staggered artefact, the so-called taste violation (analogously to ours, an alternative link-smearing scheme, the HISQ action
\cite{hisq} suppresses the taste breaking in a similar way. The latter is used
by the hotQCD collaboration in their latest studies \cite{arXiv:1111.1710}).
Taste symmetry breaking is a discretization error which is important mainly in
the low temperature phase.  In the continuum limit the physical
spectrum is fully restored. 

In analogy with what we did in Refs.~\cite{6,7}, we set the scale at the
physical point by simulating at $T=0$ with physical quark masses \cite{7} and
reproducing the kaon and pion masses and the kaon decay constant. This gives an
uncertainty of about 2\% in the scale setting. 

For details about the simulation algorithm, renormalization and a discussion
on the cut-off effects we refer the reader to \cite{7,Fodor:2009ax}.


\subsection{Finite temperature ensembles\label{sec:ensembles}} 

\begin{figure}[htb]
\begin{center}
\includegraphics[width=\textwidth]{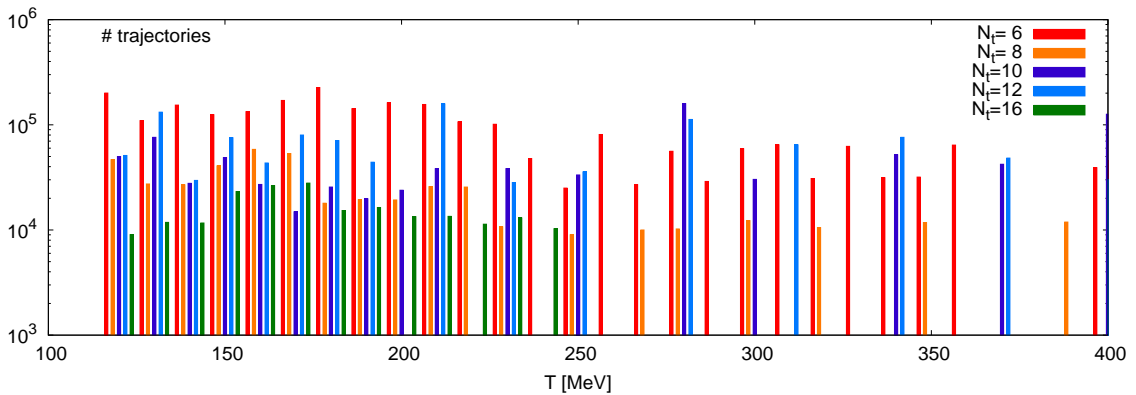}
\end{center}
\caption{\label{fig:ensembles}
The statistics used in this study. The number of trajectories 
exceeds $10^5$ for several temperatures. Each bar refers to the respective
color-coded lattice resolution in a 10 MeV wide temperature bin. We analyzed
the gauge configurations after every tenth trajectroy with 128 pairs of random
sources (256 at $N_t=16$) with the same physical quark masses that we had in
the simulation.
}
\end{figure}

The compact Euclidean spacetime of temperature $T$ and three-volume $V$ is
discretized on a hypercubic lattice with $N_t$ and $N_s$ points in the temporal
and spatial directions, respectively: 
\bea
\label{eq:t}
T = \frac{1}{N_t a},\quad\quad V = (N_s a)^3,
\eea
where $a$ is the lattice spacing. At fixed $N_t$, the temperature can be set
by varying the lattice spacing. This implies varying the bare parameters
of the lattice action accordingly, keeping the pion and kaon (Goldstone) masses
at their physical values. In other words, all our simulation points lie
on the line of constant physics, determined at zero temperature in our earlier
works \cite{6,7}. For every given $N_t$ we keep the geometry fixed, such that
the aspect ratio is $\sim 3$.

For the present analysis we use five lattice spacings for each temperature in
the transition region, corresponding to the temporal resolutions
$N_t=6,8,10,12$ and $16$. We enriched our existing set of temperatures since
Refs.~\cite{Borsanyi:2010bp,Borsanyi:2010cj} and four-folded the statistics at
$N_t=16$. We save a configuration every tenth trajectory in a hybrid monte
carlo stream with unit trajectory length. The statistics used in
this analysis is given in Fig.~\ref{fig:ensembles}, summed in bins of 10 MeV.


\subsection{Fluctuations from the lattice} 

The fluctuations of interest are derivatives of the free energy with respect
to the chemical potential of a conserved charge. This guarantees the
finiteness of the lattice observables, thus no renormalization is necessary. 

A derivative of the partition function can be written in terms of $S_{\rm
eff}$, the action with all fermionic degrees of freedom already integrated out,
as follows:
\begin{equation}
\partial_i \log Z = \frac{1}{Z} \int{\cal D}U \partial_i e^{-S_{\rm eff}} = \avr{A_i}.
\end{equation}

When we take further derivatives, the following chain rule applies:
\begin{equation}
\partial_i\partial_j \log Z=
\avr{A_iA_j}-\avr{A_i}\avr{A_j}+\delta_{ij}\avr{B_i}.
\end{equation}

Here $i$ indicates the variable of the derivative, the chemical potential $\mu_i$
in this case, with $i=u,d,$ or $s$. $A_i$ and $B_i$ are the first and second
derivatives of $S_{\rm eff}$ without the factor $e^{-S_{\rm eff}}$. Their
ensemble averages are calculated with the same weight used for generating the
configurations.

In our case $A_i$ and $B_i$ are 
\begin{eqnarray}
A_i&=&\frac14\textrm{tr}  M_i^{-1} M_i'\,,\label{eq:A}\\
B_i&=&\frac14\textrm{tr} \left(
M_i'' M_i^{-1}
-M_i' M_i^{-1} M_i' M_i^{-1}.
\right)\,.\label{eq:B}
\end{eqnarray}
with $M_i=m_i+\dslash$ is the fermion operator with the bare mass $m_i$, that
we also used for generating these configurations.  $M_i'$ and $M_i''$ stand for
its first and second derivatives with respect to $\mu_i$, respectively. 
The pre-factor $\frac14$ is required by the staggered formulation of the
single flavor trace. These derivatives are mass independent.  In the
lattice simulation as well as in the subsequent analysis, the bare mass is the
only parameter that identifies a particular type of quark. The $B$, $Q$ and $S$
quantum numbers are provided by Eq.~(\ref{chem}).

Ref. \cite{Gottlieb:1988cq} describes a stochastic technique for calculating
the traces in Eqs.~(\ref{eq:A}) and (\ref{eq:B}). The traces are rewritten
in terms of inner products of random sources. 
The most expensive part of the present analysis is the calculation of the trace
in Eq.~(\ref{eq:A}), which contains disconnected contributions and appears in
almost all susceptibilities as $\chi_{ud}$. It required 128 pairs of random
sources per configuration (256 for $N_t=16$). For each pair of sources one
needs two inversions of the fermion matrix with the light quark mass.


\subsection{Continuum extrapolation\label{sec:cont}} 

With five lattice spacings per temperature we are in the position to go
beyond the simplest form of continuum extrapolation and fit a second
order polynomial in $a^2$. Especially at low temperatures, such fit is indeed
necessary, as the coarser lattices have corrections beyond the $a^2$ term.
In general, a continuum fit benefits from higher order terms,
but this also introduces ambiguities, such as whether it is appropriate to keep
the coarsest point in the continuum extrapolation. The answer to this question
is obtained by performing all possible extrapolations, and weighting them by
the resulting goodness of the fit.  Accordingly, we varied the number
of included lattice spacings and made a linear and a quadratic fit in $a^2$.
We double the number of such choices by considering extrapolations of the
inverse fluctuations ($1/\chi$) too, and then taking the inverse of the
corresponding continuum result.

There is another source of systematic error: the interpolation ambiguity.  The
ensembles were not taken exactly at the same temperatures for different
$N_t$ values, and the spline fit on the data for a given $N_t$ depends on the
node points.  We take two choices of the node points into the analyses
(selecting the original temperature values with either even or odd indices).
In most cases the two interpolations agree within statistical errors. 
We incorporated the systematic error from this source into the statistical
error of the interpolating data points prior to the continuum fit. 
We selected the temperature range for each data set such that a consistent
interpolation is possible.

This procedure has (with few exceptions) preferred the full quartic fit 
over four or five points in the transition region, and a suppressed quartic term
fit for $T>200$ MeV. In most cases the reciprocal fit was preferred over the
original variant.

\begin{figure}[htb]
\begin{center}
\includegraphics[width=3.5in]{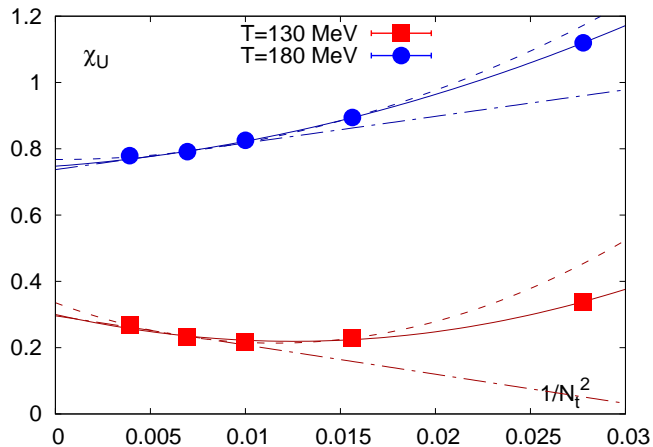}
\end{center}
\caption{\label{fig:cuextrapol}
Examples of our continuum extrapolations. Here
we show $\chi_U$, an observable severely hit both by taste breaking and by the
cut-off effects in the one-link staggered dispersion relation.  The data points
suggest a quadratic fit in $1/N_{t}^{2}$. Here we give three possible fits both
below and above the transition temperature. The solid, dashed and dash-dotted
curves represent the fits on the finest five, four and three lattices,
respectively. The uncentainty related to this choice is incorporated in the
systematic error (see main text). 
The statistical errors are much smaller than the size of the points,
nevertheless the fits provide reasonable $\chi^2$ values.
}
\end{figure}

We systematic errors are defined through the central 68.2\% of the
weighted distribution of all analyses, following our collaboration's
standard technique c.f. \cite{arXiv:0906.3599,Durr:2010aw}.
For simplicity, we give our results with the sum of the statistical
and the symmetrized systematic errors. The continuum bands in the results
section correspond to this combined error around a central value.
In most cases the systematic error dominates. For $\chi_{us}$ and $C_{BS}$ the
two types of errors are of equal magnitude.  The actual smallness of
$\chi_{ud}$ makes the relative combined error grow beyond 50\%, thus we dropped
this observable from our result list.

We give one example of our interpolation strategy in Fig.~\ref{fig:cuextrapol}.
We plot the measured $\chi_{2}^{u}$ values with statistical errors for two
different temperatures.  The data points seem to lie on a parabola, when
plotted as a function of $a^2\sim 1/N_t^2$. We give three fits for each of the
selected temperatures to indicate the spread of the possible continuum results.

As discussed in Ref. \cite{Borsanyi:2010cj}  one can use the tree level
improvement program for observables. Independently whether one used it or not
the results are the same (c.f. Figure 8 of Ref.~\cite{Borsanyi:2010cj}). For
simplicity we use in the present paper the direct method and do not apply the
tree level improvement for our observables when we extrapolate to the continuum
limit. The improvement factors for the various $N_t$
discretizations ($c_6=1.517$, $c_8=1.283$, $c_{10}=1.159$, $c_{12}=1.099$,
$c_{16}=1.054$) are merely used here for plotting the raw lattice data.



\section{Results}
\begin{figure}
\begin{minipage}{.48\textwidth}
\begin{center}
\hspace{-5cm}
\parbox{6cm}{
\scalebox{.7}{
\includegraphics{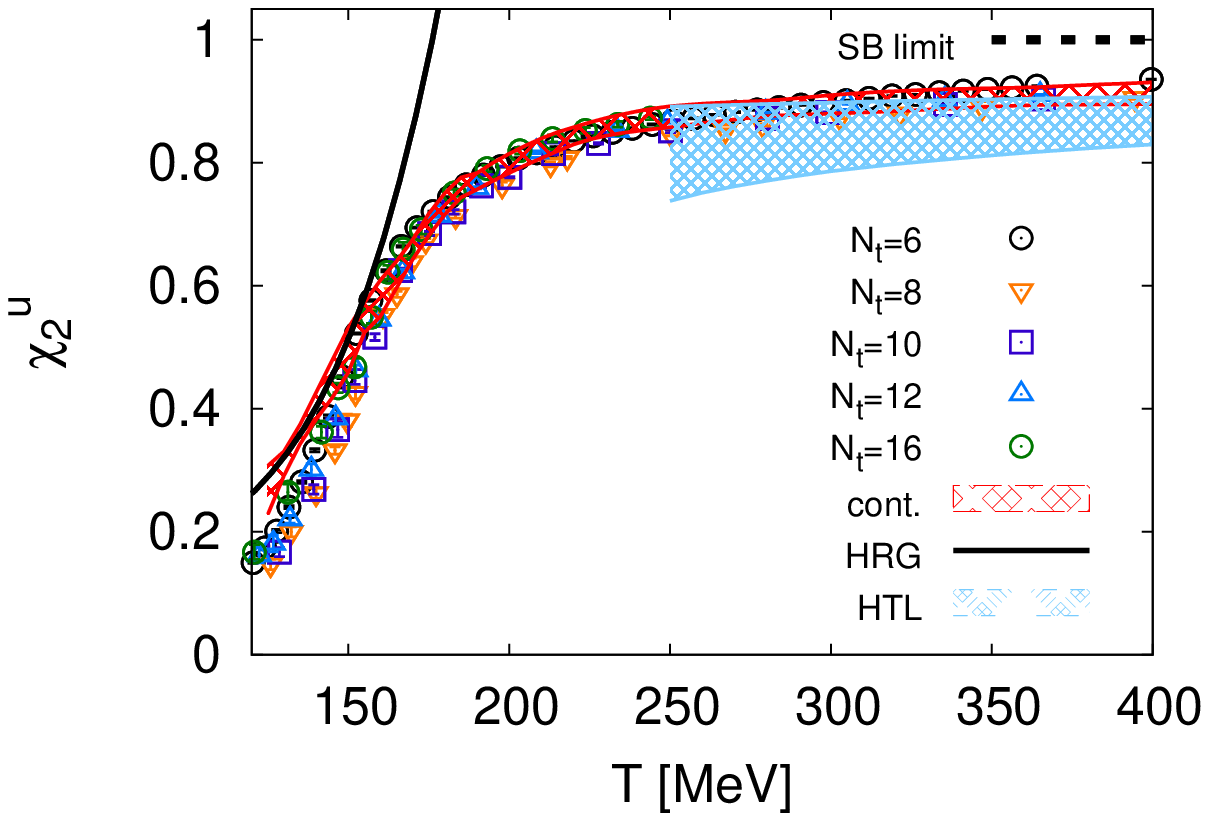}}}
\end{center}
\end{minipage}
\begin{minipage}{.48\textwidth}
\begin{center}
\hspace{-3cm}
\parbox{6cm}{
\scalebox{.7}{
\includegraphics{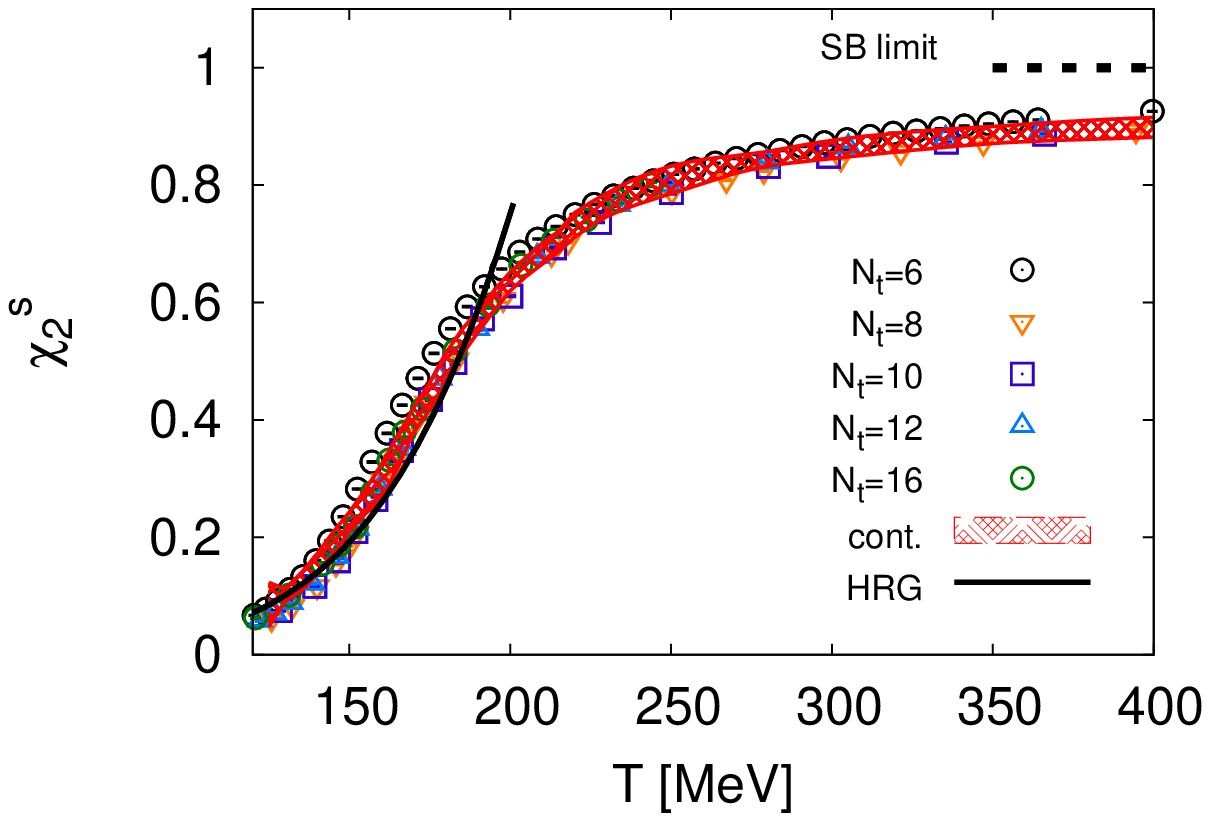}\\}}
\end{center}
\end{minipage}
\caption{Left panel: diagonal light quark susceptibility as a function of the temperature. Right panel:
diagonal strange quark susceptibility as a function of the temperature. In both panels, the different symbols correspond to different $N_t$ values. The red band is the continuum extrapolation. The black curve is the HRG model prediction for these observables. The dashed line shows the ideal gas limit. The light blue band in the left panel is the HTL prediction taken from Ref. \cite{Blaizot:2001vr}. }
\label{fig1}
\end{figure}

\begin{figure}
\begin{minipage}{.48\textwidth}
\begin{center}
\hspace{-5cm}
\parbox{6cm}{
\scalebox{.7}{
\includegraphics{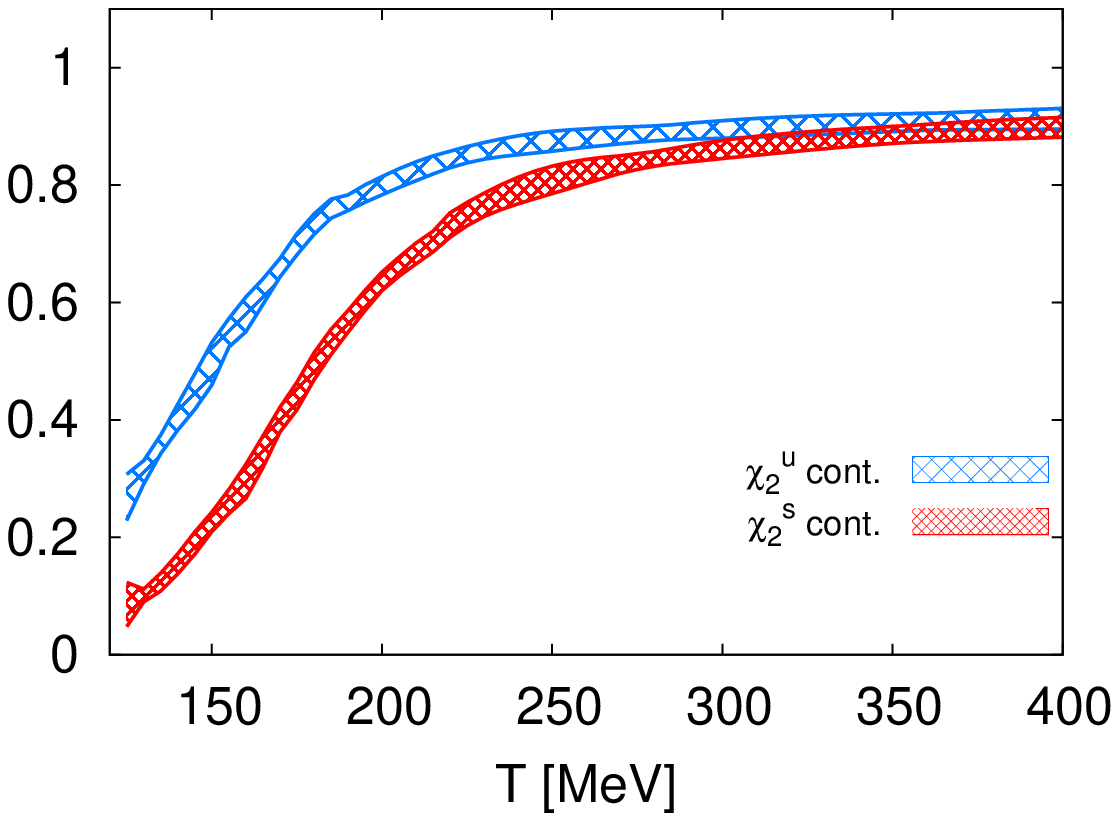}}}
\end{center}
\end{minipage}
\begin{minipage}{.48\textwidth}
\begin{center}
\hspace{-3cm}
\parbox{6cm}{
\scalebox{.7}{
\includegraphics{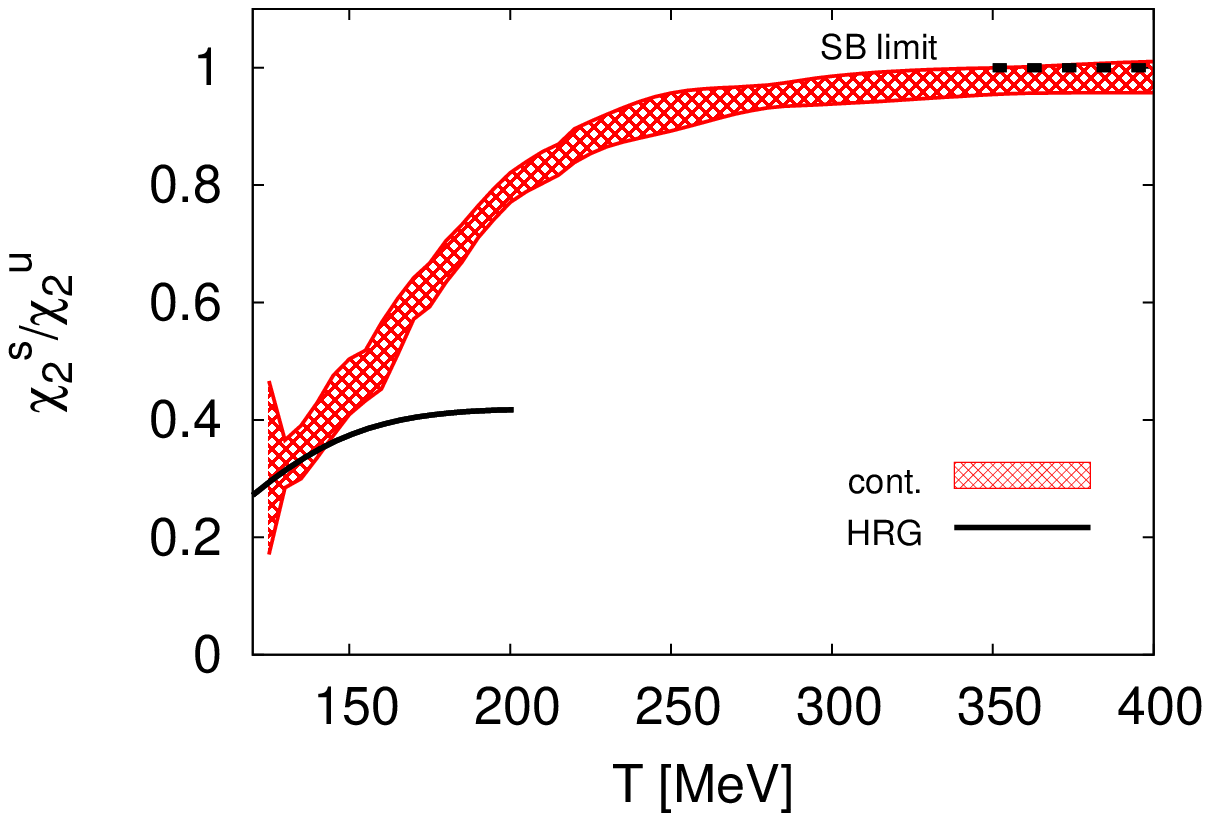}\\}}
\end{center}
\end{minipage}
\caption{Left panel: direct comparison between the continuum limit of light and strange quark susceptibilities. Right panel: ratio $\chi_{2}^s/\chi_{2}^u$ as a function of the temperature. The red band is the lattice continuum result. The black, solid curve is the HRG model prediction. The dashed line indicates the ideal gas limit.}
\label{fig2}
\end{figure}

The first observables we discuss are the diagonal light and strange quark
number susceptibilities: their behavior as functions of the temperature is
shown in the two panels of Fig. \ref{fig1}. The different symbols correspond to
different values of $N_t$, from 8 to 16. The red band is the continuum
extrapolation, obtained from the unimproved data, not from the improved ones.
The continuum extrapolation is performed through a parabolic fit in the
variable $(1/N_t)^2$, over five $N_t$ values from 6 to 16.  The band shows the
spread of the results of other possible fits, as discussed in Section
\ref{sec:cont}. The comparison between the improved data and the continuum
bands in the figure shows the success of the improvement program throughout the
entire temperature range. But even the unimproved data could be easily fitted
for a continuum limit, the combined errors are below 3\% in the deconfined phase.
Both observables show a rapid rise in a certain
temperature range, and reach approximately 90\% of the ideal gas value at large
temperatures. However, the temperature around which the susceptibilities rise
is approximately 15-20 MeV larger for strange quarks than for light quarks.
In addition, the light quark susceptibility shows a steeper rise with
temperature, compared to the strange quark one.  As expected, they approach
each other at high temperatures. 
The effect more evident in Figure \ref{fig2}: in
the left panel we show the continuum extrapolation of both susceptibilities on
the same plot. In the right
panel we show the ratio $\chi_s/\chi_u$: it reaches 1 only around 300 MeV,
while for smaller temperatures it is $<1$. It is worth noticing that all these
observables agree with the corresponding HRG model predictions for temperatures
below the transition.

The pattern of temperature dependence is strongly related the actual quark
mass. The difference between the light and strange susceptibilities here with
physical masses is more pronounced than in earlier works with not so light
pions (see E.g.  Ref.~\cite{hotQCD2}).  


\begin{figure}
\begin{center}
\hspace{-5cm}
\parbox{6cm}{
\scalebox{.85}{
\includegraphics{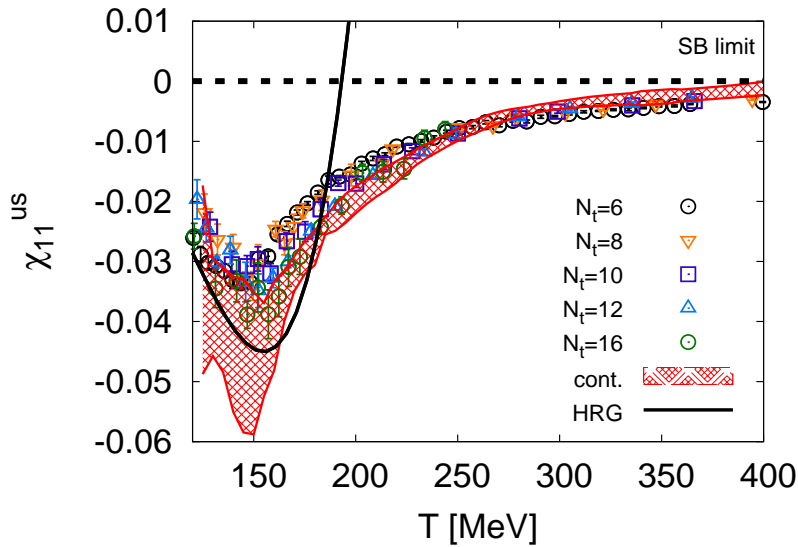}\\}}
\end{center}
\caption{Non-diagonal u-s correlator as a function of the temperature. The different symbols correspond to different $N_t$ values. The red curve is the continuum extrapolated result. The black curve is the HRG model prediction. The dashed line indicates the ideal gas limit for this observable. }
\label{fig3}
\end{figure}

The non-diagonal $us$ susceptibility measures the degree of correlation between
different flavors. This observable vanishes in the limit of an ideal,
non-interacting QGP. However, Hard Thermal and Dense Loop framework provides a
non-vanishing value for this correlation also at large temperatures
\cite{Blaizot:2001vr}.  We show our result in Fig. \ref{fig3}. $\chi_{11}^{us}$
is non-zero in the entire temperature range under study. It has a dip in the
crossover region, where the correlation between $u$ and $s$
quarks turns out to be maximal. It agrees with the HRG model prediction in the
hadronic phase. This correlation stays finite and large for a certain
temperature range above $T_c$. A
quantitative comparison between lattice results and predictions for a purely
partonic QGP state can give us information about bound
states survival above $T_c$ \cite{Ratti:2011au}.

\begin{figure}
\begin{minipage}{.48\textwidth}
\begin{center}
\hspace{-5cm}
\parbox{6cm}{
\scalebox{.7}{
\includegraphics{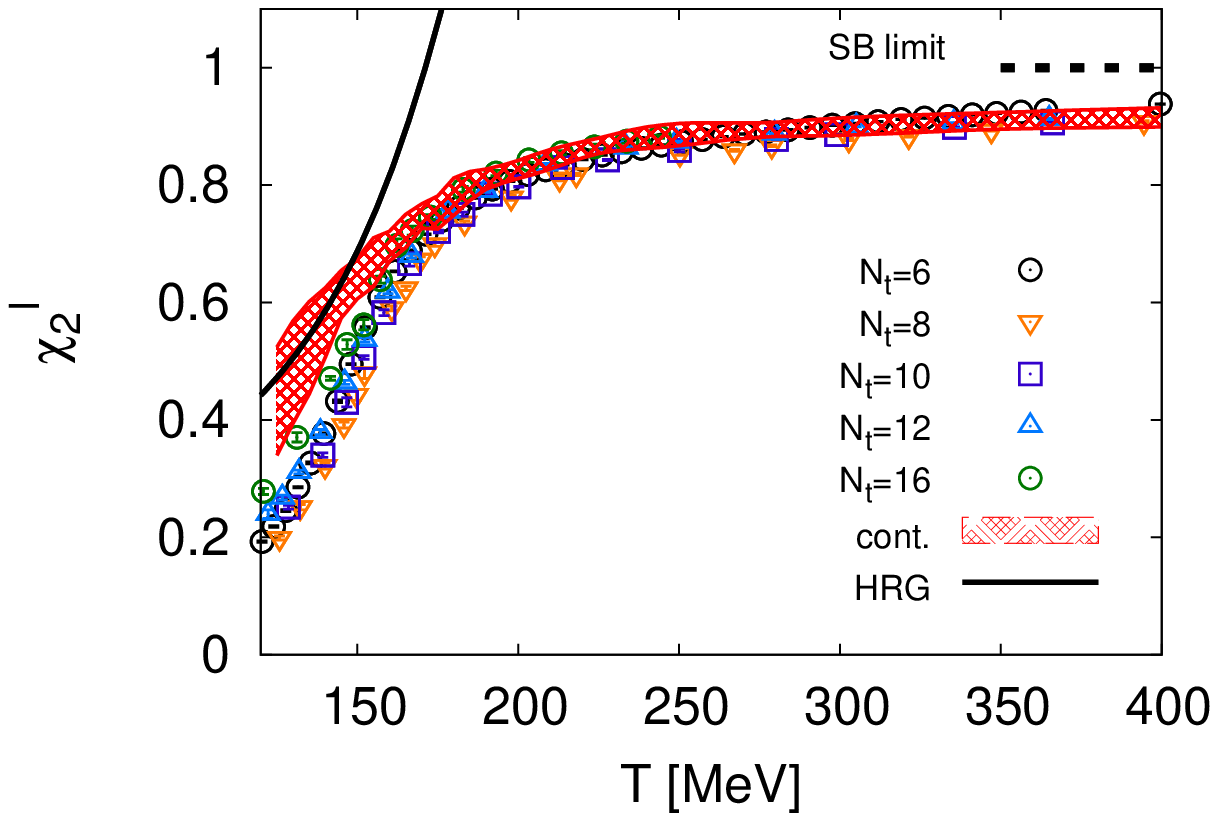}}}
\end{center}
\end{minipage}
\begin{minipage}{.48\textwidth}
\begin{center}
\hspace{-3cm}
\parbox{6cm}{
\scalebox{.7}{
\includegraphics{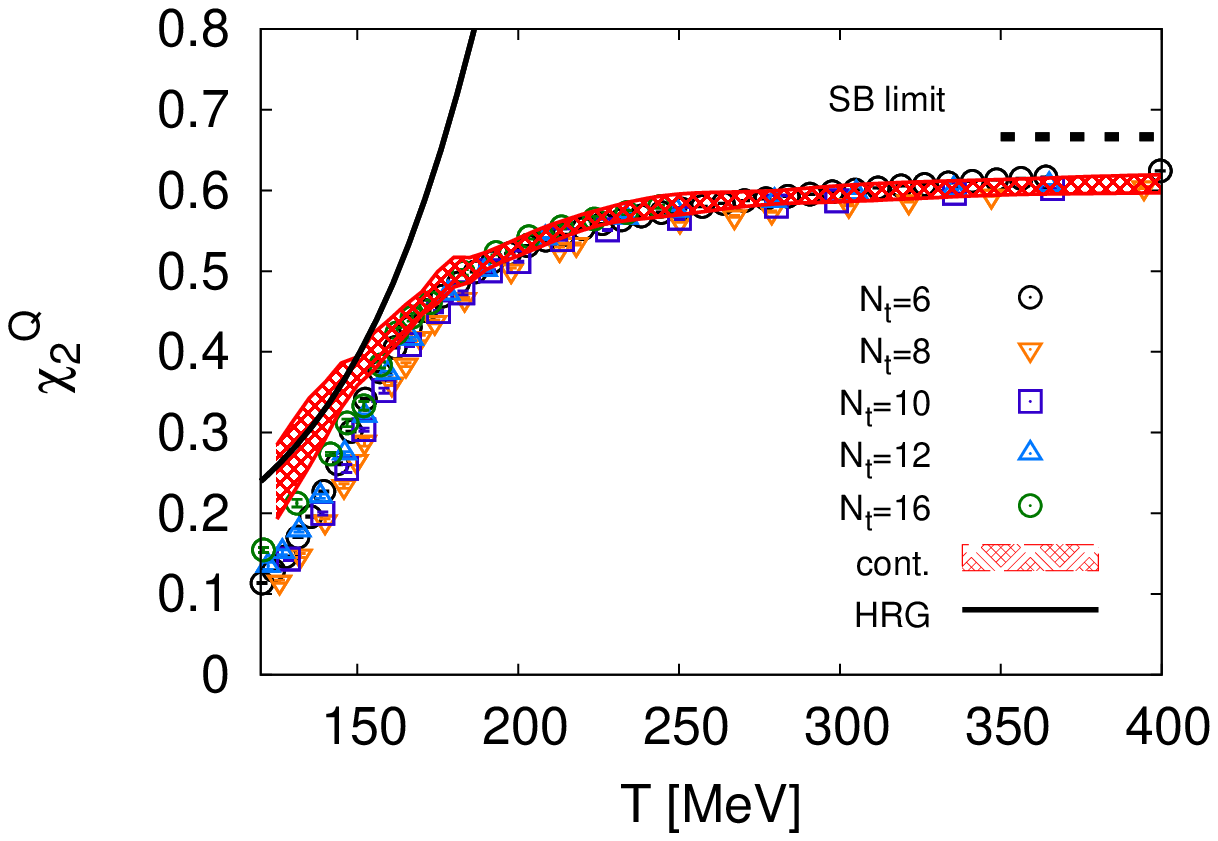}\\}}
\end{center}
\end{minipage}
\caption{Left panel: isospin susceptibility as a function of the temperature. Right panel:
electric charge susceptibility as a function of the temperature. In both panels, the different dots correspond to different $N_t$ values. The red band is the continuum extrapolation. The black curve is the HRG model prediction for these observables. The dashed line shows the ideal gas limit.}
\label{fig3a}
\end{figure}

\begin{figure}
\begin{minipage}{.48\textwidth}
\begin{center}
\hspace{-5cm}
\parbox{6cm}{
\scalebox{.7}{
\includegraphics{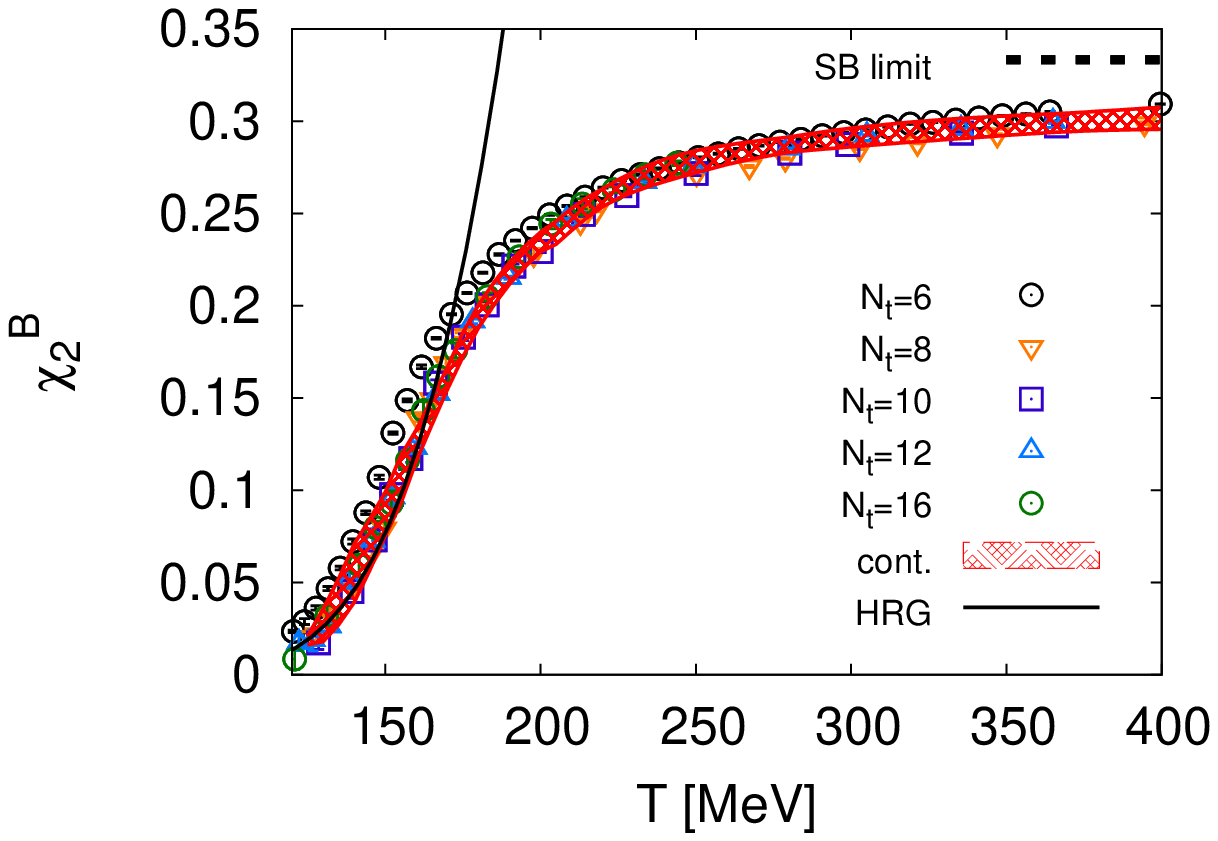}}}
\end{center}
\end{minipage}
\begin{minipage}{.48\textwidth}
\begin{center}
\hspace{-3cm}
\parbox{6cm}{
\scalebox{.7}{
\includegraphics{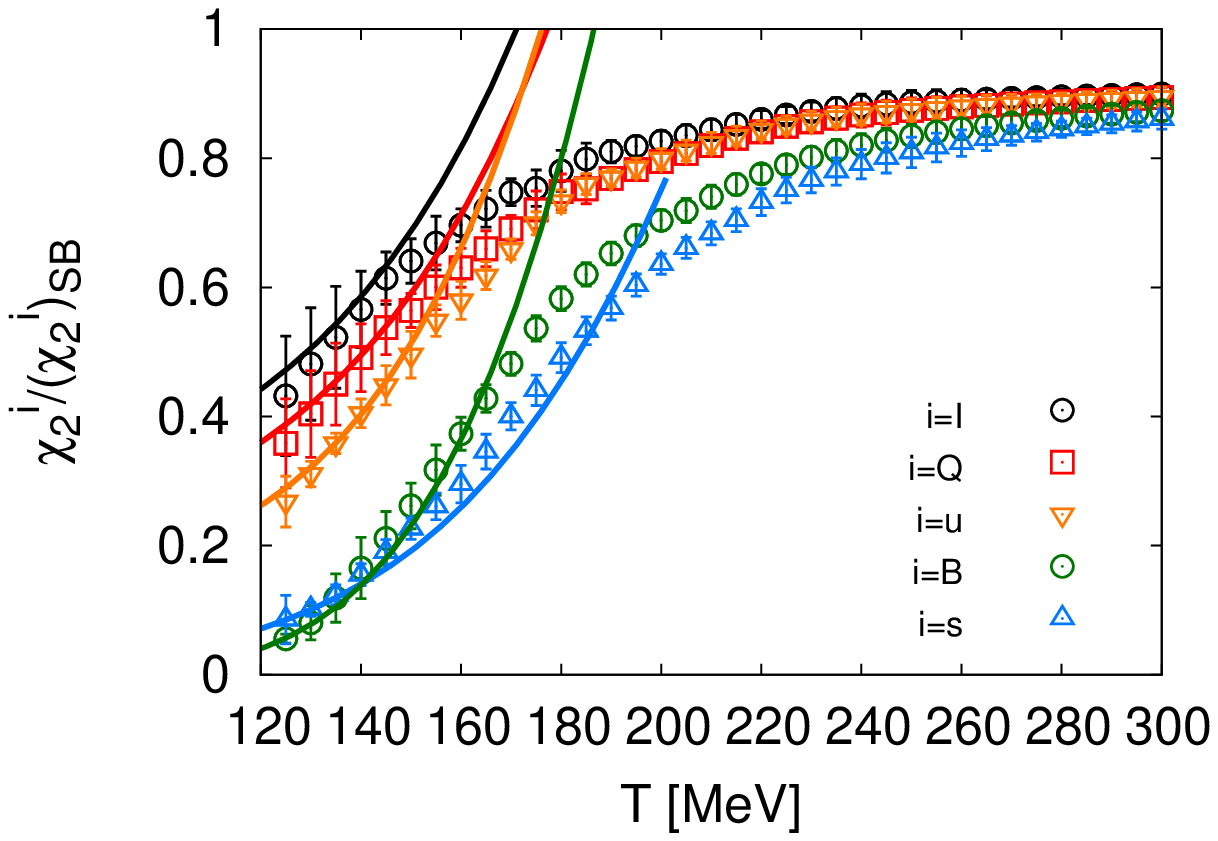}\\}}
\end{center}
\end{minipage}
\caption{Left: quadratic fluctuation of baryon number as a function of the temperature. The different symbols correspond to different $N_t$ values, the red band is the continuum extrapolation and the black, solid curve is the HRG model result. The ideal gas limit is shown by the black, dashed line.
Right: comparison between all diagonal susceptibilities, rescaled by the corresponding ideal gas limit, as functions of the temperature.}
\label{fig4}
\end{figure}

Quadratic baryon number, electric charge and isospin fluctuations can be
obtained from the above partonic susceptibilities through Eqs.~(\ref{chiB}).
We show our results for these observables in Fig.~\ref{fig3a} and in the left
panel of Fig. \ref{fig4}. In the low-temperature, hadronic phase we have a very
good agreement with the HRG model predictions. In the crossover region
these quantities all show a rapid rise with temperature, in analogy
with what already observed for the light and strange quark susceptibilities. At
large temperature they reach approximately 90\% of their respective ideal gas
values. A comparison between all diagonal susceptibilities, rescaled by their
corresponding Stefan-Boltzmann limits, is shown in the right panel of Fig.
\ref{fig4}, from which it is evident that they all show similar features in
their temperature dependence, even if the temperature at which they rise is
larger for the strangeness and baryon number susceptibilities.


\begin{figure}
\begin{center}
\hspace{-5cm}
\parbox{6cm}{
\scalebox{.85}{
\includegraphics{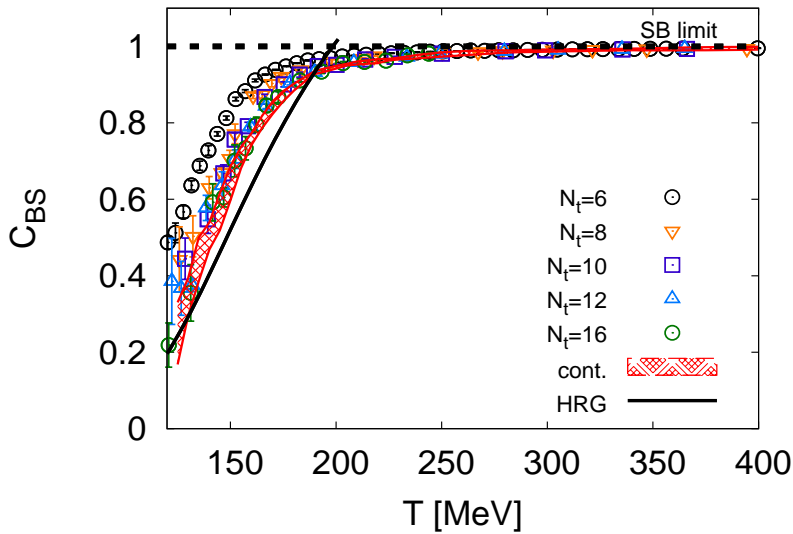}\\}}
\end{center}
\caption{ Baryon-strangeness correlator as a function of the temperature. The different symbols correspond to different $N_t$ values, the red band is the continuum extrapolation and the black, solid curve is the HRG model result. The ideal gas limit is shown by the black, dashed line.}
\label{fig6}
\end{figure}

The baryon-strangeness correlator $C_{BS}$ defined in Eq. (\ref{cbs}) was proposed long ago \cite{Koch:2005vg} as a diagnostic
for strongly interacting matter. It is supposed to be equal to one for a non-interacting QGP, while it is temperature-dependent and generally smaller than one in a hadronic system.
We show our result for this observable in Fig. \ref{fig6}.  At the smallest
temperatures it agrees with the HRG model result, and it shows a rapid rise
across the transition. It reaches the ideal gas limit much faster than
the other observables under study, yet there is a window of about 100 MeV above
$T_c$, where its value is still smaller than one. In analogy with
$\chi_{11}^{us}$, this observable also gives us information on bound state
survival above $T_c$.

For convenience we tabulate our continuum results in Table~\ref{table:result}.

\section{Conclusions}
In this paper we have presented the continuum results of our collaboration on
diagonal and non-diagonal quark number susceptibilities, in a system with 2+1
staggered dynamical quark flavors with physical masses, in a temperature range
between 125 and 400 MeV.  The continuum extrapolations were based on
$N_t=6,8,10,12$ and $16$ lattices. We calculated the systematic errors by
varying over the ambiguities of the possible extrapolations.

All observables consistently show a very good agreement with the HRG model
predictions for temperatures below the transition.

The diagonal fluctuations have some common features: they all show a rapid rise
in the crossover region, and reach approximately 90\% of the
corresponding ideal gas value at large temperatures.  The rise of both strange
quark and baryon number susceptibilities is shifted to temperatures about 20
MeV higher than those for light quark, charge and isospin susceptibilities.
Non-diagonal flavor and charge correlators remain different from their ideal
gas values for a certain window of temperatures above the transition. This
pattern encourages further studies to explore the possibility of bound state
survival above $T_c$.


\begin{table}
\hspace*{-1.5cm}\noindent\begin{minipage}{12cm}
\begin{center}
\begin{tabular}{|l|c|c|c|c|c|c|c|}
\hline
$T$ [MeV]&$\chi_2^I/T^2$&$\chi_2^U/T^2$&$\chi_2^Q/T^2$&$\chi_2^S/T^2$&$\chi_2^B/T^2$&$\chi_{11}^{us}/T^2$&$C_{BS}$\\\hline
125&0.432(92)&0.268(39)&0.239(45)&0.085(37)&0.019(2)&-0.0331(157)&0.2492(811)\\\hline
130&0.481(87)&0.311(19)&0.269(44)&0.101(10)&0.027(8)&-0.0376(80)&0.3442(492)\\\hline
135&0.523(78)&0.359(15)&0.300(42)&0.124(15)&0.040(12)&-0.0397(87)&0.4426(573)\\\hline
140&0.566(59)&0.405(22)&0.327(34)&0.155(16)&0.055(15)&-0.0438(112)&0.5040(355)\\\hline
145&0.614(40)&0.448(30)&0.358(27)&0.190(19)&0.070(13)&-0.0457(127)&0.5672(484)\\\hline
150&0.641(34)&0.496(36)&0.376(17)&0.226(16)&0.087(11)&-0.0469(118)&0.6416(442)\\\hline
155&0.669(41)&0.548(24)&0.400(22)&0.261(20)&0.106(12)&-0.0447(77)&0.7090(278)\\\hline
160&0.696(25)&0.580(29)&0.420(20)&0.295(28)&0.124(8)&-0.0407(74)&0.7575(152)\\\hline
165&0.722(28)&0.618(21)&0.440(18)&0.346(26)&0.143(7)&-0.0357(42)&0.8024(160)\\\hline
170&0.747(20)&0.659(15)&0.460(14)&0.400(21)&0.160(5)&-0.0320(28)&0.8398(163)\\\hline
175&0.753(28)&0.699(17)&0.480(19)&0.441(23)&0.179(6)&-0.0293(27)&0.8697(133)\\\hline
180&0.780(31)&0.733(17)&0.499(18)&0.491(23)&0.194(6)&-0.0277(28)&0.8931(89)\\\hline
185&0.799(24)&0.760(16)&0.502(15)&0.533(21)&0.207(6)&-0.0241(14)&0.9132(59)\\\hline
190&0.810(17)&0.770(13)&0.513(10)&0.568(18)&0.218(5)&-0.0227(21)&0.9317(38)\\\hline
200&0.827(15)&0.799(15)&0.530(10)&0.636(15)&0.235(5)&-0.0197(22)&0.9457(35)\\\hline
220&0.860(13)&0.844(14)&0.561(9)&0.732(20)&0.259(4)&-0.0151(17)&0.9616(52)\\\hline
240&0.881(18)&0.868(16)&0.578(12)&0.791(22)&0.273(6)&-0.0107(11)&0.9741(46)\\\hline
260&0.890(16)&0.880(15)&0.586(11)&0.823(20)&0.282(5)&-0.0073(10)&0.9822(54)\\\hline
280&0.895(12)&0.888(12)&0.592(8)&0.845(12)&0.287(4)&-0.0052(7)&0.9866(30)\\\hline
300&0.900(14)&0.895(15)&0.596(9)&0.861(15)&0.291(5)&-0.0040(6)&0.9889(27)\\\hline
320&0.904(15)&0.900(16)&0.600(10)&0.873(15)&0.294(5)&-0.0033(8)&0.9905(22)\\\hline
340&0.908(14)&0.905(15)&0.603(9)&0.882(14)&0.297(5)&-0.0028(11)&0.9920(20)\\\hline
360&0.911(14)&0.908(14)&0.605(9)&0.889(14)&0.299(5)&-0.0024(9)&0.9932(34)\\\hline
380&0.913(15)&0.911(15)&0.607(10)&0.894(16)&0.300(5)&-0.0018(10)&0.9943(39)\\\hline
400&0.915(16)&0.913(17)&0.608(11)&0.899(16)&0.302(5)&-0.0012(11)&0.9953(36)\\\hline
\end{tabular}

\end{center}
\end{minipage}
\caption{\label{table:result}
In this table we list the results of our continuum extrapolations.
We indicated the sum of the statistical and symmetrized systematic errors
around the central value.
}
\end{table}

\section*{Acknowledgments}
\vspace{-.2cm}
Computations were carried out at the Universities Wuppertal,  
Budapest and For\-schungs\-zentrum Juelich. This work is supported in  
part by the Deutsche Forschungsgemeinschaft grants FO 502/2 and SFB- 
TR 55 and by the EU (FP7/2007-2013)/ERC no. 208740, as well as
the Italian Ministry of Education, Universities and Research under the Firb Research Grant RBFR0814TT. 
We acknowledge the fruitful discussions with Alexei Bazavov, Christian H\"olbling,
Volker Koch and Krzysztof Redlich.

\end{document}